\documentclass[a4paper,12pt]{article}
\usepackage{amsfonts}
\usepackage{latexsym}
\usepackage{amsmath}
\usepackage{amssymb}
\usepackage{amssymb}
\usepackage{slashed}
\usepackage{upgreek}
\usepackage{mathrsfs}
\usepackage{bbm}

\usepackage{amsthm}

\theoremstyle{remark}

\usepackage[utf8]{inputenc}
\usepackage[english]{babel}

\theoremstyle{definition}

\allowdisplaybreaks

\usepackage{relsize}
\usepackage{graphicx}
\usepackage{comment}

\renewcommand{\thefootnote}{\fnsymbol{footnote}}

\def\appendix#1{\addtocounter{section}{1}\setcounter{equation}{0}
\renewcommand{\thesection}{\Alph{section}}
\section*{Appendix \thesection\protect\indent \parbox[t]{11.15cm}{#1}}
\addcontentsline{toc}{section}{Appendix \thesection\ \ \ #1}}

\font\mybb=msbm10 at 11pt

\def\bb#1{\hbox{\mybb#1}}

\def\bZ {\bb{Z}}
\def\bR {\bb{R}}

\def\bC {\bb{C}}

\newcommand{\bea}{\begin{eqnarray}}
\newcommand{\eea}{\end{eqnarray}}

\usepackage[usenames,dvipsnames,svgnames,table]{xcolor}	
\usepackage[normalem]{ulem}				
\usepackage{microtype}
\usepackage[colorlinks, citecolor=blue, linkcolor=blue, urlcolor=Maroon, filecolor=Maroon, linktocpage=true]{hyperref} 

\usepackage{chngcntr}
\counterwithout{equation}{section}

\begin{document}

\begin{center}
\vspace*{-1.0cm}
\begin{flushright}
\end{flushright}


\vspace{2.0cm} {\Large \bf Vacuum states from a resolution of the lightcone singularity  } \\[.2cm]

\vskip 2cm
 G.  Papadopoulos
\\
\vskip .6cm


\begin{small}
\textit{Department of Mathematics
\\
King's College London
\\
Strand
\\
 London WC2R 2LS, UK}\\
\texttt{george.papadopoulos@kcl.ac.uk}
\end{small}
\\*[.6cm]

\end{center}

\vskip 2.5 cm

\begin{abstract}
\noindent

The lightcone singularity at the origin is resolved by blowing up the singular point to $\bC P^1$. The Lorentz group acts on the resolved lightcone and has $\bC P^1$ as a special orbit.    Using Wigner's method of associating  unitary irreducible representations of the Poincar\'e group to particle states, we find  that the special orbit gives rise to  new vacuum states.  These vacuum states are labelled by the principal series representations of
$SL(2,\bC)$. Some remarks are included on the applications of these results to  gauge theories and asymptotically flat spacetimes.

\end{abstract}



\newpage

\renewcommand{\thefootnote}{\arabic{footnote}}



\section*{}
It is a celebrated result of Wigner \cite{wigner} that single particle states of relativistic quantum field theories are associated with unitary irreducible representations of the Poincar\'e group.  As the Poincar\'e group is the semi-direct product $SO^\uparrow(3,1)\ltimes\bR^{3,1}$, where $SO^\uparrow(3,1)$ is the proper and orthochronous  Lorentz group, its unitary irreducible representations   are constructed from the orbits of  $SO^\uparrow(3,1)$  on the dual space, $\widehat\bR^{3,1}$, of Minkowski spacetime $\bR^{3,1}$. The coordinates of  $\widehat\bR^{3,1}$ are the components of the relativistic momentum $p_\mu$, $\mu=0,1,2,3$. There are three such orbits of $SO^\uparrow(3,1)$ on  $\widehat\bR^{3,1}$ for which the time component of the momentum is positive  $p_0\geq 0$. These are the hyperboloid that lies within the future lightcone, the future part of the lightcone and the origin of $\widehat\bR^{3,1}$. The first orbit is associated with massive particle states characterised by their mass and spin, the lightcone orbit is associated with massless particles characterised by their helicity\footnote{There are additional representations in the massless case as for example the infinite or continuous  spin representations, see e.g. the review \cite{xbnb} and references within.}, and the origin of $\widehat\bR^{3,1}$ can be associated  with the vacuum state of a quantum field theory.

The lightcone, $\mathrm{LC}$, is a hyper-surface in
$\widehat\bR^{3,1}$ described by the equation $p^2=\eta^{\mu\nu} p_\mu p_\nu=0$. The $\mathrm{LC}$  is singular at $p=0$ as the rank of the differential $ \eta^{\mu\nu} p_\mu dp_\nu$ at $p=0$ is zero\footnote{View $f=p^2$ as a map from $\widehat\bR^{3,1}$ to $\bR$ and apply the inverse function theorem.}. Geometrically, the mass $m$ of a particle can be seen as a deformation parameter which can be used to resolve the  lightcone singularity of $\mathrm{LC}$ at $p=0$. Indeed the differential $ \eta^{\mu\nu} p_\mu dp_\nu$ has rank one everywhere when restricted to $p^2=m^2$, $m\not=0$.  Thus the (two sheeted) hyperboloid  $p^2=m^2$ is a smooth submanifold in $\widehat\bR^{3,1}$. The mass deformation parameter introduces an energy scale in  a theory.

 The purpose of this paper is to describe an alternative way to resolve the lightcone singularity by blowing up the singular point  and then explore some consequences that this may have in the context of quantum field theory. Different resolutions of singularities and their physical applications have been widely explored in string theory and are usually related to non-perturbative phenomena and the emergence of new states in the theory. They are also widely use in geometry, e.g. in the theory of Calabi-Yau manifolds, to resolve singularities.  Here, the resolution of the lightcone singularity that we shall pursue is the construction of a smooth space  $\widetilde {\mathrm{LC}}$ which projects on the lightcone $\mathrm{LC}$, $\pi: \widetilde {\mathrm{LC}} \rightarrow \mathrm{LC}$ such that $\pi$ is 1-1 away from the singular point $\{0\}$ and $\pi^{-1}(0)= \bC P^1$. Therefore $\widetilde {\mathrm{LC}}$ is identical to  $\mathrm{LC}$ away from the singular point and the singular point of $\mathrm{LC}$ is blown up in $\widetilde {\mathrm{LC}}$ to $\bC P^1$. As for the mass deformation, the above resolution of the lightcone singularity also introduces a scale in a theory set by the radius of $\bC P^1$.

To explore applications to quantum field theory, we shall use the Wigner's method to identify states with orbits of the Lorentz group in  $\widetilde {\mathrm{LC}}$. The blow up construction leads to a natural action of the Lorentz group on $\widetilde {\mathrm{LC}}$. In particular, the Lorentz group has two orbits on
$\widetilde {\mathrm{LC}}_+=\pi^{-1}({\mathrm{LC}}_+) $, where $ {\mathrm{LC}}_+$ is  the future lightcone including the singularity at $\{0\}$. One orbit, the principal orbit,  is the usual orbit of the Lorentz group on the $ {\mathrm{LC}}_+-\{0\}$ - after all $\widetilde {\mathrm{LC}}_+-\pi^{-1}(0)$ and $ {\mathrm{LC}}_+-\{0\}$ are identical.  Another orbit, the special orbit,
is $\pi^{-1}(0)= \bC P^1$.  The latter admits a natural action of $SL(2,\bC)/\bZ_2$, where $SL(2,\bC)$ is  the spin group of $SO^\uparrow(3,1)$. The application of the Wigner's method on
$\widetilde {\mathrm{LC}}_+-\pi^{-1}(0)$ yields the usual massless representations of the Poincar\'e group. However there are additional representations
arising from the action of $SL(2,\bC)$ on the special orbit $\bC P^1$.  These irreducible unitary representations are associated with the principal series of $SL(2,\bC)$ and they are labelled by $[k, \nu]$, where $k\in \bZ/2$ and $\nu\in \bR$.  These new states are identified with vacua. As we shall demonstrate the Poincar\'e group acts with the trivial representation as in the case of the standard vacuum of a relativistic quantum field theory.

 First let us describe $\widetilde {\mathrm{LC}}$. There is a general theory of how one can blow up points on manifolds but in the case of the  $\mathrm{LC}$ in $\widehat\bR^{3,1}$, the construction takes a  simple form.  To begin consider the space $\widehat\bR^{3,1}\times \bC P^1$ and impose the algebraic equation
 \bea
 \big( p_0 {\bf 1} +\vec p \cdot \vec \sigma\big)u=0
 \label{blc}
 \eea
 where ${\bf 1}$ is the identity $2\times 2$ matrix, $\vec \sigma$ are the Pauli matrices and  $u\in \bC P^1$.  $u$ in this context can be thought of as a non-vanishing column vector with components the complex numbers $z_1$ and $z_2$ defined up to an overall scale by a complex number. The equation (\ref{blc}) is the chiral Dirac equation written in momentum  space.

The space $\widetilde {\mathrm{LC}}$ is defined as   the space of  solutions of  (\ref{blc}) in  $\bR^{3,1}\times \bC P^1$. For $p\not=0$, $\widetilde {\mathrm{LC}}$ is identified with ${\mathrm{LC}}$.  To see this observe that (\ref{blc}) implies $p^2u=0$ and as $u\not=0$, one has that $p^2=0$, i.e. (\ref{blc}) implies the lightcone condition. For $p\not=0$, (\ref{blc}) has a unique solution $u$ up to an overall scale, i.e. the solutions of
 (\ref{blc}) are labelled by the points of ${\mathrm{LC}}$ as $u\in \bC P^1$ is determined uniquely for each $p$.  On the other hand for $p=0$, the solutions of (\ref{blc}) are all the points of $\bC P^1$, i.e. the singular point $p=0$ of ${\mathrm{LC}}$ has been replaced with $\bC P^1$ in $\widetilde {\mathrm{LC}}$.  One can check that $\widetilde {\mathrm{LC}}$ is a smooth submanifold of $\bR^{3,1}\times \bC P^1$ as a consequence of the inverse function theorem.  $\widetilde {\mathrm{LC}}$ is a smooth resolution of ${\mathrm{LC}}$, where the lightcone singularity has been replaced with $\bC P^1$.

 Let us describe $\widetilde {\mathrm{LC}}$ in more detail.   For this rewrite (\ref{blc}) as
 \bea
 \begin{pmatrix}p_+& \bar q
 \\
 q & p_-\end{pmatrix}\begin{pmatrix} z_1\\ z_2\end{pmatrix}=0
 \eea
 where $p_\pm=p_0\pm p_3$ and $q=p_1+ip_2$. First suppose that $z_1\not=0$. In such a case one finds that
 \bea
 p^{(1)}_+= p_-^{(1)} |\zeta^{(1)}|^2~,~~~q^{(1)}=-p^{(1)}_- \zeta^{(1)}
 \label{p1}
 \eea
 where $\zeta^{(1)}=z_2/z_1$ and the superscripts indicate the patch for which $z_1\not=0$.  Therefore the coordinates of this patch are $\{p_-^{(1)}, \zeta^{(1)}, \bar\zeta^{(1)}\}$.

 Next suppose that $z_2\not=0$ and the solution of (\ref{blc}) can be written as
 \bea
 p_-^{(2)}=p_+^{(2)} |\zeta^{(2)}|^2~,~~~q^{(2)}=-p^{(2)}_+ \zeta^{(2)}~,
 \label{p2}
 \eea
 where $\zeta^{(2)}=z_1/z_2$, i.e. the coordinates of this patch are $\{p_+^{(2)}, \zeta^{(2)}, \bar\zeta^{(2)}\}$.  Clearly $\widetilde {\mathrm{LC}}$ is covered by these two patches and the patching condition  is
\bea
\zeta^{(1)}={1\over \zeta^{(2)}}~,~~~p_-^{(1)}=p_+^{(2)} |\zeta^{(2)}|^2
\label{pat}
\eea
where the first equation  is the familiar patching condition of $\bC P^1$. 

To see how $SO^\uparrow(3,1)$ acts on $\widetilde {\mathrm{LC}}$ first observe that $\widehat\bR^{3,1}\times \bC P^1$ admits an action of $SO^\uparrow(3,1)\times SL(2,\bC)$, where $SL(2,\bC)$ is not necessarily identified with the spin group of $SO^\uparrow(3,1)$.  $SL(2,\bC)$  acts on
$\bC P^1$ with the familiar fractional linear transformations which is the group of holomorphic diffeomorphisms or equivalently conformal transformations of $\bC P^1$. This action descends to an action of $SL(2, \bC)/\bZ_2=SO^\uparrow(3,1)$ on $\bC P^1$ as the elements $A$ and $-A$ of $SL(2,\bC)$ act with the same transformation on $\bC P^1$. It is clear that everywhere that $p\not=0$, $SO^\uparrow(3,1)\times SL(2,\bC)/\bZ_2$ leaves the algebraic equation (\ref{blc}) invariant   provided that $SL(2,\bC)$ is identified with the spin group of $SO^\uparrow(3,1)$. As a consequence of this identification,  $p\not=0$ solutions of (\ref{blc}) transform to solutions  of (\ref{blc}). For $p=0$, $SO^\uparrow(3,1)$ acts trivially on the solutions of (\ref{blc}).  But the group\footnote{ Although $SL(2,\bC)/\bZ_2$ is isomorphic to $SO^\uparrow(3,1)$, we retain the $SL(2,\bC)/\bZ_2$ notation to distinguish that action of $SL(2,\bC)$ on $\bC P^1$ from that of the Lorentz group on the rest of the space.} $SL(2,\bC)/\bZ_2$ still acts non-trivially  on the solutions of (\ref{blc})   with the fractional linear transformations.  Continuity  suggests that $SL(2,\bC)$ should still be identified with the spin group of $SO^\uparrow(3,1)$.

In particular on the coordinates $(p^{(1)}_-, \zeta^{(1)}, \bar \zeta^{(1)})$  and $(p^{(2)}_+, \zeta^{(2)}, \bar \zeta^{(2)})$ of $\widetilde {\mathrm{LC}}$, $SO^\uparrow(3,1)$ acts as
\bea
\tilde\zeta^{(1)}={c+d \zeta^{(1)}\over a+ b \zeta^{(1)}}~,~~~\tilde p^{(1)}_-= p^{(1)}_- |a+b \zeta^{(1)}|^2~,
\label{tran1}
\eea
and
\bea
\tilde\zeta^{(2)}={a\zeta^{(2)} +b \over c \zeta^{(2)}+ d }~,~~~\tilde p^{(2)}_+= p^{(2)}_+ |d+c \zeta^{(2)}|^2~,
\label{tran2}
\eea
where
\bea
\begin{pmatrix} a & b\\ c& d\end{pmatrix}\in SL(2,\bC)~.
\eea
Observe that the above group action is compatible with the patching conditions (\ref{pat}), i.e. the blow up operation is equivariant with respect to the group action of $SO^\uparrow(3,1)$ on the lightcone.

The group $SO^\uparrow(3,1)$ acting as described above on $\widetilde {\mathrm{LC}}_+$ has two orbits for $p_0\geq 0$.  One is the familiar $\widetilde {\mathrm{LC}}_+-\pi^{-1}(0)$ orbit associated with the massless representations of the Poincar\'e group and the other is $\pi^{-1}(0)=\bC P^1$.  Note that the condition $p_0\geq 0$ is consistent with both the group action of the Lorentz group on $\widetilde{\mathrm LC}$ and the patching conditions (\ref{pat}) as $p_0^{(1)}=p_0^{(2)}$.

Before we proceed further let us give a few more details on the construction of unitary representations  of $SO^\uparrow(3,1)\ltimes\bR^{3,1}$.  As it has been already mentioned, the unitary irreducible representations of $SO^\uparrow(3,1)\ltimes \bR^{3,1}$ are characterised by the orbits of $SO^\uparrow(3,1)$ acting on the characters of the unitary irreducible representations of $\bR^{3,1}$.  These characters are maps $\chi: \bR^{3,1}\rightarrow \bC$ such  that $\chi(a_1+a_2) =\chi(a_1) \chi(a_2)$  and can be parameterised  with the momentum  $p$ as $\chi_p(a)= e^{i\langle p, a\rangle}$,  $p \in \widehat\bR^{3,1}$, where   $\langle\cdot,\cdot\rangle$ is the natural pairing of $\bR^{3,1}$ with  its dual space $\widehat\bR^{3,1}$.  Then for each orbit one considers the induced representations  of $SO^\uparrow(3,1)$; for the general theory of induced representations see \cite{wigner, mackey1, mackey2, isham, warner}.  These are unitary representations  labelled by a representations $\rho$ of the isotropy group in $SO^\uparrow(3,1)$ of the  orbit.  Given such an induced  representation $U$ of $SO^\uparrow(3,1)$, one can construct an irreducible unitary representation of $SO^\uparrow(3,1)\ltimes\bR^{3,1}$ by setting
\bea
\Big({\mathcal U}\big((\Lambda, a)\big) \psi\Big)(p)=e^{i \langle p,a\rangle} \Big(U\big(\Lambda\big) \psi\Big)(p)~,
\label{tranact}
\eea
 where $(\Lambda, a)\in SO^\uparrow(3,1)\ltimes\bR^{3,1}$, $p$ is in one of the  orbits mentioned above  and  $\psi$ is a section of a suitable vector bundle over the orbit specified with $\rho$. For example, the isotropy group of $\mathrm{LC}_+-\{0\}$   is $SO(2)\ltimes\bR^2$ and $\rho$ is one of the finite dimensional irreducible representations of $SO(2)$ associated with the helicity of the state.

 Focusing on the orbits of $SO^\uparrow(3,1)$ on $\widetilde {{\mathrm LC}}_+$, the representation of  $SO^\uparrow(3,1)\ltimes\bR^{3,1}$ induced by  the generic orbit $\widetilde {{\mathrm LC}}_+-\pi^{-1}(0)$ is identified with that on $\mathrm{LC}_+-\{0\}$  associated with massless particles and described above.  As $p=0$ for the special orbit $\pi^{-1}(0)=\bC P^1$, $SO^\uparrow(3,1)$ acts trivially on the orbit as it can be seen either from $p\rightarrow p \Lambda $ or equivalently from (\ref{tran1}), (\ref{tran2}), (\ref{p1}) and (\ref{p2}). The same applies for translations as it can be seen from (\ref{tranact}) for $p=0$. However $SL(2, \bC)/\bZ_2$  acts non-trivially on $\pi^{-1}(0)=\bC P^1$.  It is well known that $\bC P^1$ induces the principal series of unitary irreducible representations of $SL(2,\bC)$. The isotropy group is given by the matrices of the form
 \bea
 \begin{pmatrix} \ell& w\\ 0& \ell^{-1}\end{pmatrix}
 \eea
 where $\ell\in \bC-\{0\}$ and $w\in \bC$. The representation of the little group used is
 \bea
 \rho_{\nu, k} \begin{pmatrix} \ell& w\\ 0& \ell^{-1}\end{pmatrix}= r^{i\nu} e^{2ik\theta}
 \eea
 where $\ell=r e^{i\theta}$, $k\in \bZ/2$ and $\nu\in \bR$. It turns out that the representations $k$ and $-k$ give isomorphic induced representations.
 These are new states  which are associated with the special orbit $\pi^{-1}(0)$ and they are labelled with $\psi_{\nu, k}$. They are sections of complex line bundles over $\bC P^1$.  As we have seen, they are invariant under $SO^\uparrow(3,1)\ltimes\bR^{3,1}$ and they are localized on $\pi^{-1}(0)$. Because of their invariance under the action of $SO^\uparrow(3,1)\ltimes\bR^{3,1}$, they are identified with new  vacuum states of a theory. $k$ can be thought as the helicity of the states as the representations of the same $SO(2)$ subgroup of $SL(2,\bC)$ label the helicity in the standard massless representations of $SO^\uparrow(3,1)\ltimes\bR^{3,1}$.

  Processes in QED and gravity are infrared finite even though individual diagrams can be infrared divergent \cite{low, gell, kazes, yennie, weinberg1, weinberg2}.  However  non-abelian gauge theories in four dimensions and in particular QCD are infrared divergent in perturbation theory. In the infrared these theories are strongly coupled with quarks and gluons confined to mesons and hadrons. It has been suggested that there is chiral symmetry breaking with the light mesons identified as the goldstone bosons. There is also an explicit chiral symmetry breaking at lower energies with the scale set by the masses of the lightest quarks.
   Our analysis provides some evidence that in theories with massless particles there is a  new intricate vacuum structure that arises in the infrared.
This is reminiscent of the work on  asymptotically  flat spacetimes  reviewed in \cite{strominger}, see also references within.
 There are  some similarities in the construction.  For example, the  expression for the momentum   (\ref{p2}) in one of the patches in terms of the complex coordinates $(\zeta, \bar \zeta)$ is the same as that of the equation (2.8.13) in \cite{strominger} up to  notation.  However there is no mention of our patching condition $\zeta\rightarrow 1/\zeta$ dictated by the solution of the Dirac equation (\ref{blc}) which in standard coordinates is a reflection $x_1,x_2\rightarrow -x_1, -x_2$ and  $x_3\rightarrow x_3$ and leads to a smooth space.  The gluing  condition $\zeta\rightarrow -1/ \bar\zeta$   which is significant in the scattering processes considered in \cite{strominger} is   the antipodal map $\vec x\rightarrow -\vec x$ on the 2-sphere.  This can be incorporated in our description by considering the anti-chiral Dirac equation
 \bea
 \big( p_0 {\bf 1} -\vec p \cdot \vec \sigma\big)v=0~,
 \label{blc1}
 \eea
 where again $v\in \bC P^1$. Then notice that if $u$ is a solution of (\ref{blc}), then $v=i \sigma_2 \bar u$ will be a solution of (\ref{blc1}). It is straightforward to notice that the transformation $i\sigma_2 *$ acting on $\bC P^1$ is the antipodal map  $\zeta\rightarrow -1/ \bar\zeta$, where $*$ is the complex conjugation operation $*u=\bar u$.  Both formalisms rely on the existence of a non-shrinking $\bC P^1$ as the energy of the massless fields goes to zero. In our case this $\bC P^1$ is the blow up of the singular point of the lightcone while in \cite{strominger} is the past (future) boundary\footnote{The $\mathcal{I}^+_-$ should not be identified with the spatial infinity $i^0$ of the conformal boundary as in most conformal compactifications of asymptotically flat  black holes and that of Minkowski spacetime $i^0$ is a point and not a 2-sphere.}   $\mathcal{I}^+_-$  ($\mathcal{I}^-_+$) of the  future null infinity $\mathcal{I}^+$ ($\mathcal{I}^-$). Also a large part of the  construction in \cite{strominger} is in configuration space (spacetime) while our construction is in momentum space. Nevertheless despite the differences and the different starting points, the two constructions have many similarities and may be connected.

     It has been proposed in e.g. \cite{sachs, newman} that for asymptotically flat spacetimes the BMS group should replace the Poincar\'e group.  As a result the unitary representations of the  BMS group \cite{bms1, bms2}  have been extensively investigated and classified in \cite{mc1}-\cite{ap1}. This is rather delicate as one has to investigate the dual space of the super-translations group which depends on the topology that one puts on the group. One choice is to identify  the group of super-translations with  the space of square integrable functions on $S^2$.  This is  a Hilbert space and so isomorphic to its dual.  The orbits of $SL(2,\bC)$ on the space of square integrable functions on $S^2$ have been investigated in \cite{mc2} and has been found  that  they have compact isotropy groups.  As a result  the massless representations of $SO^\uparrow(3,1)\ltimes\bR^{3,1}$  do not lift to unitary irreducible representations of the BMS group. Alternatively one can choose the nuclear topology on the BMS group. This is sufficient to show that both the massive and  massless representations of $SO^\uparrow(3,1)\ltimes\bR^{3,1}$ lift to unitary irreducible representations of the BMS group \cite{mc4}  but still this is not the case with the representations of the principal series of $SL(2, \bC)$. Therefore if $SL(2,\bC)/\bZ_2$ of the BMS group is identified with the $SL(2,\bC)/\bZ_2$ group of $\tilde {\mathrm{LC}}$, the vacuum states to not admit an action of the BMS group. Alternatively  if $SL(2,\bC)/\bZ_2$ of the BMS group is identified with the Lorentz group $SO^\uparrow(3,1)$ acting on momentum $p$ in the standard way, $SL(2,\bC)/\bZ_2$ as well as the subgroup of spacetime translations of the supertranslations group  act on the new vacuum states with the trivial representation. Further work is needed to establish how the remaining generators of the group of supertranslations act on the new vacuum states.
     In addition note the BMS group has many more additional unitary representations which do not have an apparent physical application.

The unitary irreducible representations of the Poincar\'e group have been investigated acting on  a resolution of the lightcone which replaces the lightcone singularity at the origin with $\bC P^1$. We have found that apart from the usual  massless representations of the Poincar\'e group, there are additional representations which can be identified with those of the principal series of the Lorentz group. It has been argued that these representations lead  to new vacuum states in the theories with massless particles. It is with some hesitation that  physical applications of the above mathematical result have been explored.   Nevertheless, the resolution of the lightcone singularity as its (mass) deformation is rather natural and compelling  and because of this it have been presented.

\section*{Acknowledgments}

I would like to thank Ulf Gran for helpful discussions and for reading the manuscript, and  Dionysis Anninos for discussions on the unitary representations of non-compact groups.


\end{document}